# Tunable superconductivity in Fe-pnictide heterointerfaces by diffusion control

Silvia Haindl[1,*], Sergey Nikolaev[1], Michiko Sato[2], Masato Sasase[2], Ian MacLaren[3]

[1] Tokyo Tech World Research Hub Initiative (WRHI), Institute of Innovative Research, Tokyo Institute of Technology, 4259 Nagatsuta-cho, Midori-ku, Yokohama, Kanagawa 226-8503, Japan

[2] Materials Research Center for Element Strategy, Tokyo Institute of Technology, 4259 Nagatsuta-cho, Midori-ku, Yokohama, Kanagawa 226-8503, Japan

[3] School of Physics and Astronomy, University of Glasgow, Glasgow G12 8QQ, UK

* corresponding author's e-mail address: s.haindl.aa@m.titech.ac.jp

New Fe-pnictide heterostructures of the type *Ln*OFeAs/BaFe$_2$As$_2$ (*Ln* = La, Sm) were grown by pulsed laser deposition (PLD) and investigated. Their common structural unit of [Fe$_2$As$_2$] planes allows perfect matching between the different Fe-pnictide unit cells and a coherent and atomically sharp interface. We test the stability of the heterointerface in the presence of Co$^{2+}$ (cations) as well as for excess O$^{2-}$ (anions) and discuss the consequences on the electronic properties: While undoped SmOFeAs/BaFe$_2$As$_2$ remains non-superconducting, a balanced Co-concentration after diffusion across the interface results in superconductivity within Co-substituted variants. In contrast, excess O$^{2-}$ causes the formation of an interfacial layer in SmOFeAs/BaFe$_2$As$_2$ with increased O$^{2-}$/As$^{3-}$ ratio and develops a metal-to-superconductor transition with time. The engineered heterointerfaces may provide a sophisticated pathway to bridge the gap between Fe-pnictides and Fe-chalcogenides.

The monolayer (*ML*-) FeSe/SrTiO$_3$ heterointerface between an Fe-chalcogenide and a perovskite oxide has recently demonstrated an ability to host high-temperature superconductivity in the range of 40 – 75 K [1 – 3], although the origin of Cooper pairing is still controversial [4, 5]. As van der Waals compounds, FeSe and FeTe have also been employed in the formation of heterointerfaces with topological insulators in the search for Majorana bound states [6,7]. While interface engineering in *ML*-FeSe/SrTiO$_3$ has attracted worldwide attention, the realization of equivalent experiments for Fe-pnictides is currently less developed and highly challenging due to their polar surfaces.

Examples of heterointerfaces for Fe-pnictides can be found in the growth of Ba(Fe$_{1-x}$Co$_x$)$_2$As$_2$/Fe [8, 9], BaFe$_2$As$_2$/SrTiO$_3$ [10 – 13] and Ba(Fe$_{1-x}$Co$_x$)$_2$As$_2$/BaFe$_2$As$_2$:O [14]. Following the widely accepted typology of non-polar and polar surfaces and interfaces, respectively, which is based on electrostatic arguments and has been successfully used in the description of heterointerfaces such as Ge/GaAs and LaAlO$_3$/SrTiO$_3$ [15 – 19], an assessment of heterointerfaces within Fe-pnictides is given in Fig. 1.

Here, we focus on heterointerfaces (Fig. 1e), that develop between *Ln*OFeAs, with *Ln* = La, Sm (*Ln*-1111), having ZrCuSiAs-type structure and the ternary Fe-pnictide, BaFe$_2$As$_2$ (Ba-122), that adopts a ThCr$_2$Si$_2$-type structure. Both, ternary and quaternary Fe-pnictides share common features, such as [Fe$_2$As$_2$]$^{2-}$ layers and ionic bonding along the crystallographic [001] direction. Uncompensated dipole moments are responsible for reconstruction and charge modulation on a BaFe$_2$As$_2$ surface [20], and may also lead to the observed Fermi surface shifts in BaFe$_2$As$_2$ after surface doping [21].

Like Ba-122/Ba-122 (Fig. 1d), the *Ln*-1111/Ba-122 interface does not interrupt the sequential stacking of layers with alternating layer charge ±2 and thus avoids a polarization discontinuity (Fig. 1f). We prove below that the commonly shared [Fe$_2$As$_2$]$^{2-}$ layer constitutes a nonpolar (stable) heterointerface after joining *Ln*-1111 and Ba-122 unit cells with atomic precision as designed in Fig.1e. Besides the recently emphasized *geometric design plans* [13], we turn the attention towards *electrostatic principles* with their vital consequences in the engineering of films and interfaces. Both, coherency and stability of the Fe-pnictide heterointerfaces can be disturbed by changes in the local charge distributions and the ionic environment. We have, therefore, tested the interface stability in the presence of $Co^{2+}$ supply and excess $O^{2-}$ by selecting different target compositions for $LnO_{1-y}Fe_{1-x}Co_xAs$ ($Ln$ = La, Sm and $x$ = 0, 0.15; $y$ = 0, 0.2) after $BaFe_2As_2$ deposition. We show results of five heterostructures investigated by X-ray diffraction and reflectivity (XRD, XRR), scanning transmission electron microscopy (STEM), electron energy loss spectroscopy (EELS), electron-dispersive spectroscopy (EDS), Auger electron spectroscopy (AES) and electrical transport measurements. Basic information of the heterostructures and band structure calculations based on density functional theory (DFT) can be found in the Supplementary Information (SI, Tab. S1 and Fig. S1).

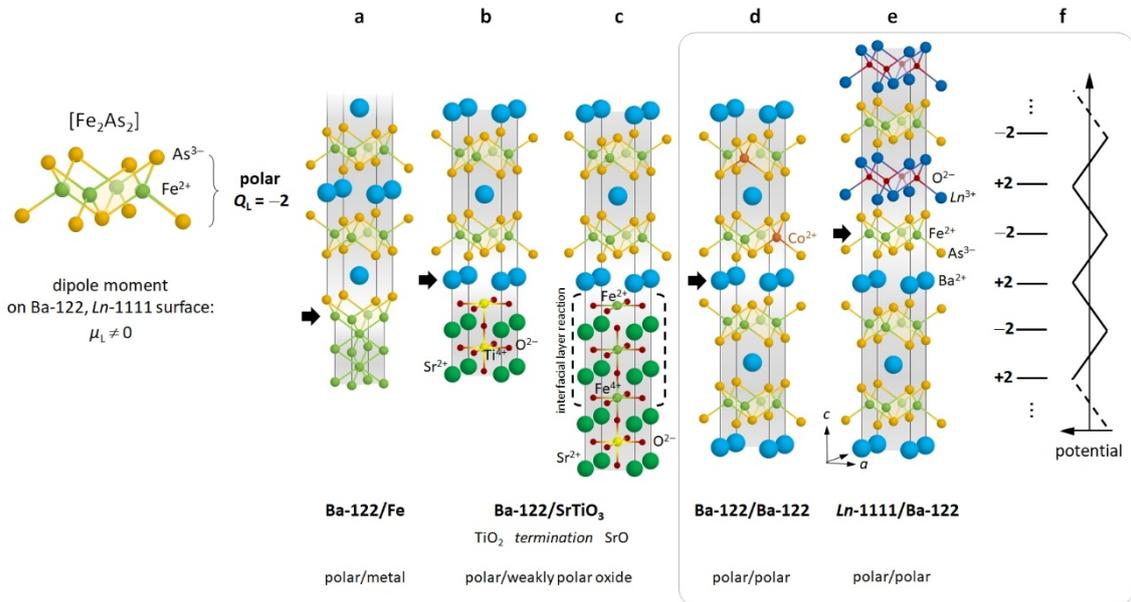

Figure 1. Fe-pnictide ($BaFe_2As_2$) heterointerfaces. The structural unit of [$Fe_2As_2$] layers has a net charge of $Q_L$ = -2 and a dipole moment, $\mu_L \neq 0$. a) Polar/metal interface in Ba-122/Fe located in the Fe plane of the [$Fe_2As_2$] layer (black arrow) [8], b) polar/weakly polar oxide interface in Ba-122/$SrTiO_3$ for $TiO_2$ termination with a shared Ba/Sr plane [12] and c) for SrO termination with the formation of an interfacial $BaFeO_{2+x}$ layer (represented here with reduced thickness) [12]. When both layers are Fe-pnictides, there is the possibility for d) a charge compensated polar/polar heterointerface between two different Ba-122 layers proposed in Refs. [14, 33] and e) a charge compensated polar/polar heterointerface between $Ln$-1111 and Ba-122 with a shared [$Fe_2As_2$] layer (this work). Although both layers are polar, they share the same sequence of layer charges resulting in a non-polar interface. f) The expected converging electric potential across Fe-pnictide/Fe-pnictide heterostructures.

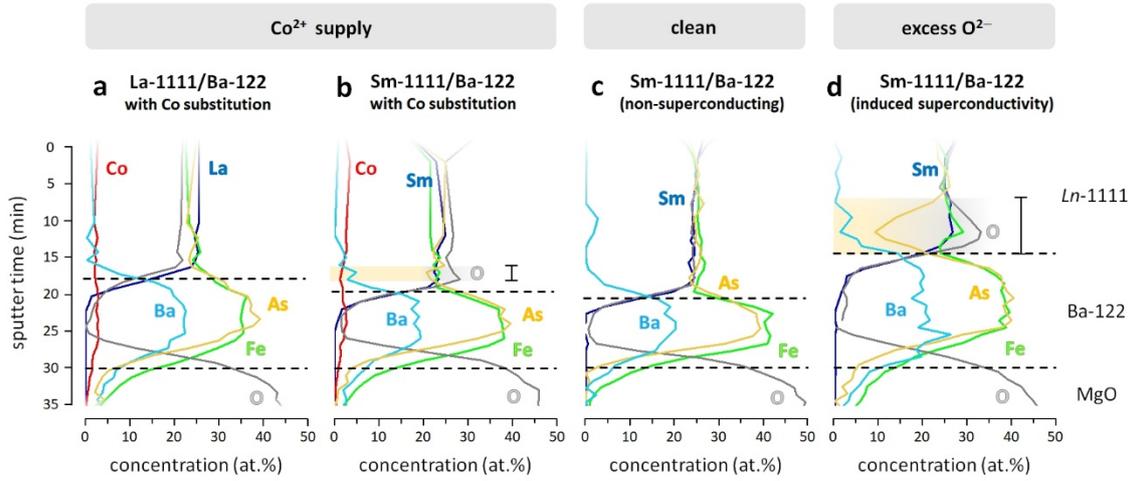

Figure 2. Summary of *Ln*-1111/Ba-122 concentration profiles measured by AES. From left to right: Superconducting, Co-substituted *Ln*-1111/Ba-122 heterostructures for a) *Ln* = La and b) *Ln* = Sm; c) a non-superconducting, clean Sm-1111/Ba-122 heterostructure; d) a Sm-1111/Ba-122 heterostructure with excess $O^{2-}$ for which an induced metal-to-superconductor transition was found. Interfacial layers with increased O-content and simultaneous As-deficiency are indicated by a vertical bar.

**Structure and composition of Fe-pnictide heterointerfaces**

We start with an overview of chemical composition profiles for the different *Ln*-1111/Ba-122 heterostructures (Fig. 2), where important observations are made in the comparison between a stoichiometric (clean) Sm-1111/Ba-122 heterostructure (Fig. 2c), two Co-substituted variants (Fig. 2a,b) and a Sm-1111/Ba-122 heterostructure with excess $O^{2-}$ (Fig. 2d).

$Co^{2+}$ ions, supplied during *Ln*-1111 deposition, do not perturb any layer or interface polarity and diffuse from the top *Ln*-1111 into the bottom Ba-122 layer down to the substrate (Fig. 2a,b). Co-diffusion leads to electron doping of the originally undoped Ba-122 layer and the final Co concentration depends on the Co supply during film growth (i.e. the composition of the $LnOFe_{1-x}Co_xAs$ target) and the layer thicknesses. In the relevant heterostructures we have used $LnOFe_{1-x}Co_xAs$ targets with $x = 0.15$. The actual average Co-content in $LnOFe_{1-x}Co_xAs/Ba(Fe_{1-x}Co_x)_2As_2$ is $x = 0.08 \pm 0.02$ (in La-1111/Ba-122) and $0.06 \pm 0.03$ (in Sm-1111/Ba-122) with an almost balanced distribution along the cross section. As a consequence, 3D superconductivity develops in the complete heterostructure, which we will discuss below in more detail.

Upon the supply of excess $O^{2-}$ the AES depth profile revealed significant differences (Fig. 2d) compared to the clean Sm-1111/Ba-122 heterostructure (Fig. 2c). The presence of excess $O^{2-}$ leads to the formation of an interfacial layer with increased $O^{2-}/As^{3-}$ ratio of up to 3.7. The interfacial layer formation can be understood in terms of a perturbed interface polarity when excess $O^{2-}$ is present. The information obtained by AES is, furthermore, consistent with the qualitative differences found in X-ray reflectivity when comparing both Sm-1111/Ba-122 heterostructures: In the presence of excess $O^{2-}$ the measured XRR intensity oscillations could only be modeled when an additional interfacial layer with larger interface roughness was introduced (SI, Fig. S6).

Microstructural investigations (HAADF-STEM) of a La-1111/Ba-122 heterostructure resolved the lattice planes of both Fe-pnictides and revealed the anticipated coherent interface with a commonly shared [$Fe_2As_2$] layer (Fig. 3a-c). This result agrees fully with the assumption of a stable interface between two polar layers with identical dipoles. The atomic resolution of the HAADF-STEM images allows the determination of the distance between successive [$Fe_2As_2$] layers, $d_{Fe-Fe\|c}$, and $c$-axis lattice parameters across the interface and in its vicinity (Fig. 3d). The distance between the last Ba plane in Ba-122 and the first La plane in La-1111 is ~6.65 ± 0.28 Å. The $c$-axis lattice parameters determined by XRD reveal $c_{1111,XRD}$ = 8.76 Å (La-1111) and ½$c_{122,XRD}$ = 6.47 Å (Ba-122), which are close to the interfacial distances from optimized DFT calculations (8.790 Å; 6.574 Å). The translational symmetry breaking along $c$-axis direction includes the exchange of Ba-planes by [$La_2O_2$] layers. This transition between the tetragonal space groups I4/mmm (no. 139) and P4/nmm (no. 129) can be viewed as the loss of an additional *in-plane* translation of (0.5 0.5 0) with the stacking of the filling layers between [$Fe_2As_2$].

The interfacial matching strongly affects the lattice parameters of the last row of Ba-122 and the first row of La-1111 unit cells. Both unit cells share a common $a$-axis lattice parameter of $a_I \approx$ 4.22 Å and are additionally elongated in $c$-axis direction with $c_{122,I} \approx 13.0 \pm 0.2$ Å (from $d_{Ba-Ba}$) and $c_{1111,I} \approx 9.25 \pm 0.55$ Å (from $d_{La-La}$), where the subscript I denotes 'interface'. Compared to the averaged $c$-axis lattice parameters from XRD, the locally determined interfacial $c$-axis lattice parameters are larger by 0.5% (for Ba-122) and larger by up to 5.5% (for La-1111). The interface itself is characterized by an abrupt step in $c$-axis lattice parameters. The separation of Fe planes, $d_{Fe-Fe,\|c}$, jumps from ~6.8 Å to ~9.5 Å when going from Ba-122 to La-1111.

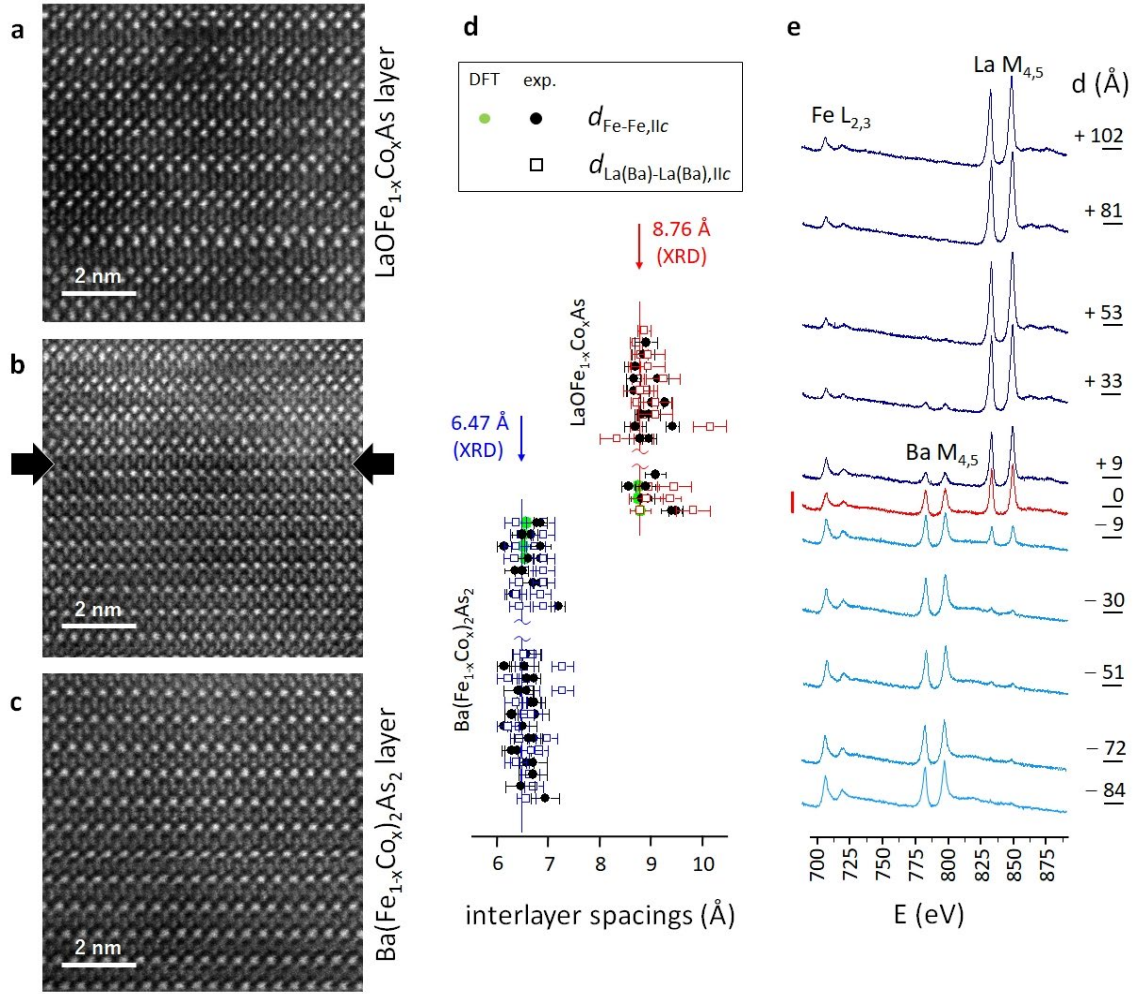

Figure 3. a) – c) HAADF-STEM of LaOFe$_{1-x}$Co$_x$As/Ba(Fe$_{1-x}$Co$_x$)$_2$As$_2$ lamella prepared by conventional milling showing from top to down the LaOFe$_{1-x}$Co$_x$As layer, a coherent interface with the shared [Fe$_2$As$_2$] layer (black arrows), and the bottom Ba(Fe$_{1-x}$Co$_x$)$_2$As$_2$ layer. The electron beam is parallel to the [100] direction of the Fe-pnictides. d) Locally (STEM) determined interlayer spacings across the heterostructure compared with globally (XRD) determined lattice parameters. DFT calculated values of the interfacial unit cell are shown as green full circles. $d_{Fe-Fe,\|c}$ jumps by 2.9 ± 0.5 Å at the interface (SI Figs. S5, S7c). e) Electron energy-loss (EELS) spectra for an energy loss range $E$ = 688.0 eV – 892.7 eV. The change in intensity of Fe L$_{2,3}$, Ba M$_{4,5}$, and La M$_{4,5}$ edges is mapped for different positions across the La-1111/Ba-122 interface (red) and traces a locally confined La$^{3+}$ ↔ Ba$^{2+}$ interdiffusion. The Co L$_{2,3}$ edges (very close to the Ba M$_{4,5}$ edges) could not be resolved in this measurement. More details are given in SI, Tab. S3.

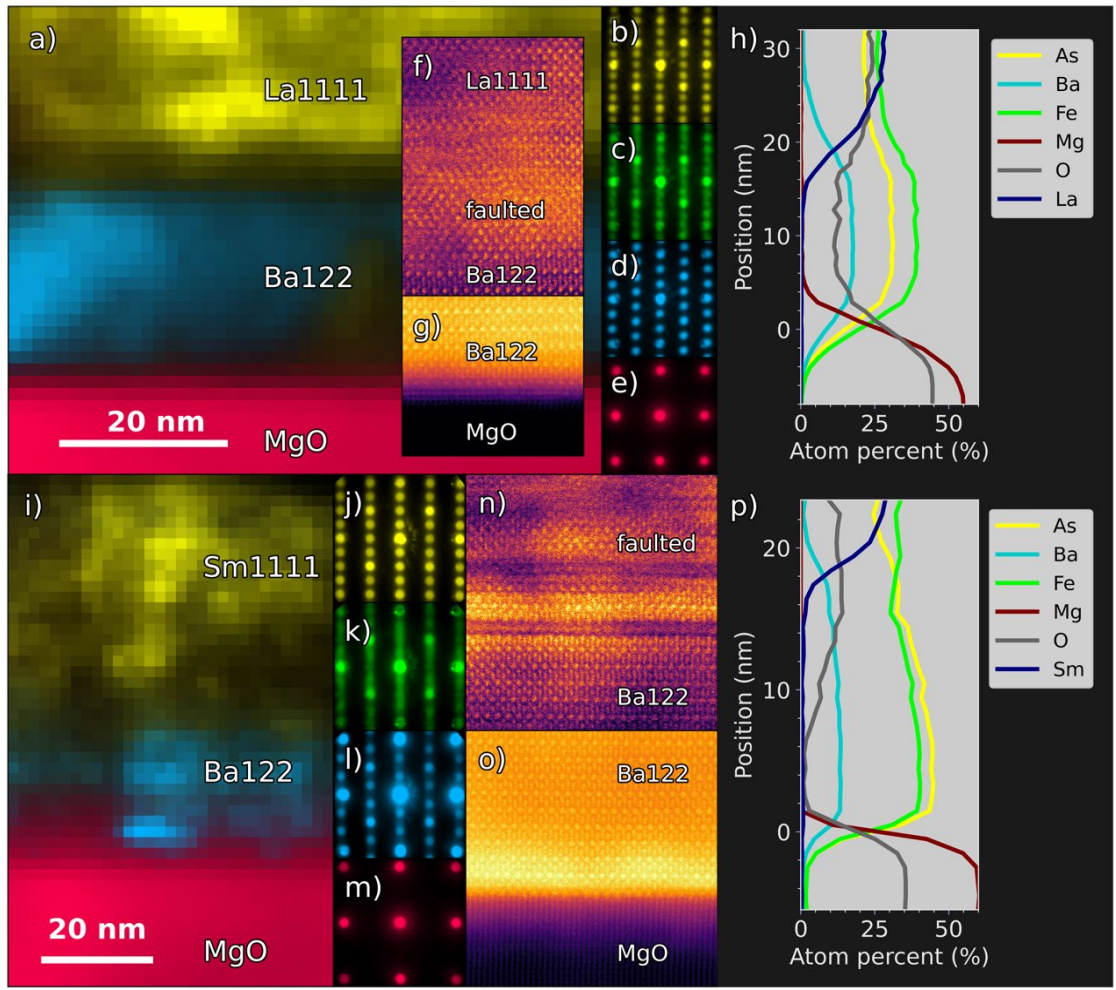

Figure 4. Scanning TEM investigations of the a-g) LaOFe$_{1-x}$Co$_x$As/Ba(Fe$_{1-x}$Co$_x$)$_2$As$_2$/MgO and h-n) SmOFe$_{1-x}$Co$_x$As/Ba(Fe$_{1-x}$Co$_x$)$_2$As$_2$/MgO samples: a) virtual dark field images for the La-1111, Ba-122 and MgO structures combined into a three-color image; b)-e) representative diffraction patterns for La-1111, faulted region, Ba-122 and MgO; f) HAADF-STEM image of the La-1111/Ba-122 interface showing some faulting and unclear structure; g) HAADF-STEM image of the Ba-122/MgO interface showing a sharp and well-defined interface; h) elemental composition traces across one such area (not exactly the same as in the images) in the same orientation as the images; i) virtual dark field images for the Sm-1111, Ba-122 and MgO structures combined into a three-color image j)-m) representative diffraction patterns for Sm-1111, faulted region, Ba-122 and MgO; n) HAADF-STEM image of the Sm-1111/Ba-122 interface showing both clearly defined faults and some unclear structure; o) HAADF-STEM image of the Ba-122/MgO interface showing a sharp and well-defined interface; p) elemental composition traces across one such area (not exactly the same as in the images) in the same orientation as the image.

Employing EELS we have traced the cation interdiffusion of Ba$^{2+}$ and La$^{3+}$ ions at least 5 nm deep into the adjacent layers (Fig. 3e). Assuming charge neutrality, the diffusion of two $Ln^{3+}$ ions should compensate for the diffusion of three $Ba^{2+}$ ions in the opposite direction. The larger ionic radius of $Ba^{2+}$ compared to $La^{3+}$ could partially explain an increase of interfacial unit cell volumes, however, while La substitution in $BaFe_2As_2$ is documented, the structural stability of LaOFeAs is expected to be strongly limited when $Ba^{2+}$ substitutes for $La^{3+}$. The presence of $Co^{2+}$ does not change the electrostatic considerations and is thus not detrimental to the Fe-pnictide heterointerface.

Lanthanide substitution at Ba sites in Ba-122 could result in a locally confined interfacial electron doping since superconductivity was reported for $(Ba_{1-x}La_x)Fe_2As_2$ films [22], whereas Sm-substitution in Ba-122 remained unachieved in previous film growth attempts [23]. At present, we can only determine that the Sm gradient into Ba-122 is similar to that of La when comparing AES depth profiles (Fig. 2). We anticipate that we do not find superconductivity in a clean Sm-1111/Ba-122 heterostructure, i.e. without $Co^{2+}$ nor excess $O^{2-}$ (SI, Fig. S1).

Another $LaOFe_{1-x}Co_xAs/Ba(Fe_{1-x}Co_x)_2As_2$ heterostructure and a $SmOFe_{1-x}Co_xAs/Ba(Fe_{1-x}Co_x)_2As_2$ heterostructure were analyzed by both scanning precession electron diffraction (SPED) and by HAADF-STEM combined with an EDS mapping across their interfaces. Fig. 4 shows the results of this analysis. The SPED data reveals the distinct diffraction patterns found in the different layers and the resulting four dimensional dataset of two real space and two reciprocal space dimensions [24] can then be processed to show the spatial distribution of each exemplar diffraction pattern [25]. These show characteristically that there are three clearly diffraction patterns for both the heterostructures, one for the MgO, one for the Ba-122 structure and one for the *Ln*-1111 structure. These are all different enough that the spots can be isolated and the spatial distribution of these three patterns can be easily mapped as virtual dark field (VDF) images. Additionally, there is often a region at the top of the Ba-122 layer which is faulted and which produces a complex diffraction pattern that appears to resemble a mixture of the 122 and the 1111 diffraction patterns, as such, this cannot be mapped as a separate pattern. Figs. 4a and 4g show three-colour composite images made from VDF images of the MgO, 122 and 1111 structures. Figs. 4b-e and 4h-k show exemplar diffraction patterns for the different layers. Additionally, insets are included of HAADF STEM images of the two interfaces in each case. The Ba-122/MgO interface is typically perfect and coherent and shows few defects (although there is a darker region in the center of Fig. 4a where there was a growth defect with a less clear structure). On the other hand, the *Ln*-1111/Ba-122 interface typically has some faulted regions around it where the structure is less clear and may not be the same through the thickness of the lamella examined in the TEM. The EDS traces in a direction perpendicular to the interfaces are compared in Figs. 4h and 4p. The EDS traces do not contain Co, as the Co-K$\alpha$ line could not be separated from the much stronger Fe-K$\beta$ and the Co-K$\beta$ was too weak to use. The absolute quantification results in EDS should be treated with caution as the quantification was not based on standards. In particular, O quantification is difficult as the O-K peak is at low energy with other peaks close by – this would be better done by EELS in future. In either case, however, it does not seem the interfaces are chemically sharp, and there does seem to be some $O^{2-}$ in the 122 layer, at least according to EDS.

We point out here, that the presence of $O^{2-}$ in Ba-122 is expected to cause structural changes, because the incorporation of a chalcogen ion into Ba-122 (either at a lattice or an interstitial site) leads to a charge imbalance across the layered Fe-pnictide, for which the ion valences at the individual crystallographic sites are largely fixed. The same can be expected for an incorporation of $O^{2-}$ in *Ln*-1111.

$O^{2-}$ is naturally supplied during growth of the *Ln*-1111 layer and can easily enter the ~20 nm thin Ba-122 layer when it does not possess a dense structure, exemplarily shown by an AFM image of the surface of bare Ba-122 film (SI, Fig. S2). Other sources of $O^{2-}$ include the Ba-122 target itself, that incorporates small amounts of $O^{2-}$ during ageing even when stored in the desiccator. We did not find any evidence that $O^{2-}$ may diffuse after film growth through the artificially designed grain boundary between *Ln*-1111 and Ba-122, as we repeated an AES analysis on one heterostructure after six months and could not detect any changes in the element concentrations. Any interlayer formation is, therefore, completed after film growth. O-diffusion from the MgO substrate can be ruled out.

**Superconductivity in Fe-pnictide heterointerfaces**

Electrical transport measurements revealed superconductivity in *Ln*-1111/Ba-122 heterostructures with $Co^{2+}$ and excess $O^{2-}$, but neither for a clean Sm-1111/Ba-122 heterostructure nor in films of BaFe$_2$As$_2$ film (20 nm) on MgO(100) or in LnOFe$_{1-x}$Co$_x$As films on MgO(100), even in the presence of a doping agent like Co-substitution. This result was attributed to strain [26]. Our DFT calculations for the La-1111/Ba-122 heterointerface (SI, Fig. S7) with an atomic arrangement such as experimentally observed in Fig. 3b predicts a smooth change in the band structure and the Fermi surface projections along [001].

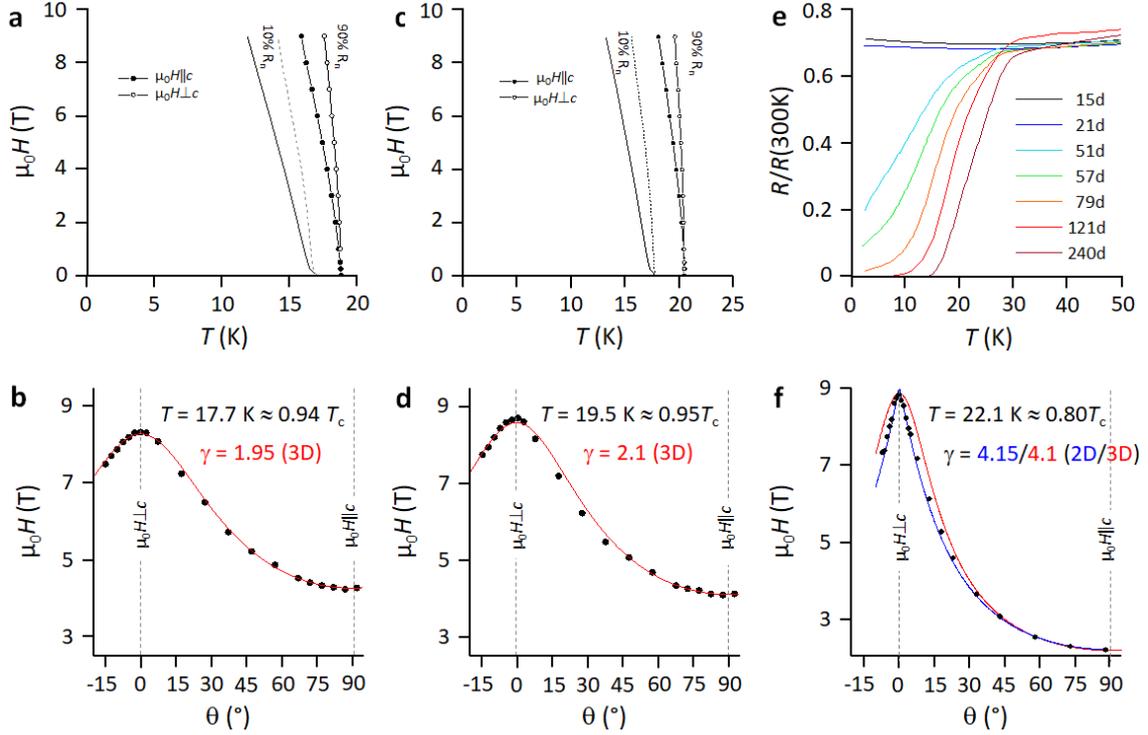

Figure 5. Superconducting properties of Ln-1111/Ba-122 heterostructures. a) Upper critical fields, $\mu_0 H_{c2}$ (II$c$ and $\perp c$) for Co-substituted La-1111/Ba-122 up to 9 T and b) angular dependence of $\mu_0 H_{c2}$ (evaluated for a 90% criterion at $T$ = 17.7 K). The red line is a fit according to AGL theory. c) Upper critical fields, $\mu_0 H_{c2}$ (II$c$ and $\perp c$) for Co-substituted Sm-1111/Ba-122 up to 9 T and d) angular dependence of $\mu_0 H_{c2}$ (evaluated for a 90% criterion at $T$ = 19.5 K). The red line is a fit according to AGL theory. e) Induced metal-to-superconductor transition in Sm-1111/Ba-122 with excess $O^{2-}$ (up to 240 days after growth) and f) angular dependence of $\mu_0 H_{c2}$ (evaluated for a 50% criterion at $T$ = 22.1 K 240 days after growth). Red curve fits are according to AGL theory, the blue curve fit is according to Tinkham's 2D formula.

In the presence of Co$^{2+}$ the measured superconducting transition temperature exceeds the maximum $T_c$ found for LnOFe$_{1-x}$Co$_x$As compounds, that is 13 – 18 K (for Ln = La – Sm). This result is easily understood in terms of Co-diffusion from the LnOFe$_{1-x}$Co$_x$As layer into the BaFe$_2$As$_2$ layer. The diffusion process during film growth balances the Co-content in the individual Fe-pnictide layers: The Co-content in the top $Ln$OFe$_{1-x}$Co$_x$As layer decreases while the Co-content increases towards optimal doping in the initially undoped Ba-122 at the bottom. The average Co-contents, $x$ = 0.08 ± 0.02 (in La-1111/Ba-122) and 0.06 ± 0.03 (in Sm-1111/Ba-122) are close to optimal doping in Ba-122, which is consistent with the observed $T_{c,90}$ of 16.5 – 20.5 K (Fig. 5 a,c). Since superconductivity develops in the whole heterostructure, a 3-dimensional (3D) superconducting state is traced by the angular dependence of the upper critical field (Fig. 5 b,d), $\mu_0 H_{c2}(\theta) = \frac{\mu_0 H_{c2\perp c}}{\sqrt{cos^2\theta + \gamma^2 sin^2\theta}}$. Minimum and the maximum of the upper critical field correspond to $\mu_0 H_{c2\parallel c}$ and $\mu_0 H_{c2\perp c}$. In the frame of the anisotropic Ginzburg-Landau (AGL) theory, the $H_{c2}$-anisotropy $\gamma$ can be expressed as anisotropy of the electronic mass, $\gamma = \frac{H_{c2\perp c}}{H_{c2\parallel c}} = \sqrt{\frac{m_{\parallel c}}{m_{\perp c}}}$. The Co-substituted heterostructures have a low $H_{c2}$-anisotropy of $\gamma \approx 2.0 \pm 0.1$ at temperatures close to $T_c$.

Sm-1111/Ba-122 heterostructures with excess $O^{2-}$ and an interfacial layer formation develop an induced metal-to-superconductor transition over time (Fig. 5e, SI, Fig. S1). We attribute this induced superconductivity to electronic/structural modifications caused in the interfacial layer with an increased $O^{2-}/As^{3-}$ ratio. The as-grown heterostructure is initially non-superconducting. A superconducting transition emerges after several weeks with $T_c$ increasing, as displayed in Fig. 5e, up to 240 days after growth showing $T_{c,90}$ = 27.5 K and a complete superconducting transition. Compared to the Co-substituted variants, the larger $T_{c,90}$ indicates less disorder at the Fe-sites but sufficient doping. The angular dependence of $\mu_0 H_{c2}$ deviates from a 3D AGL fit with an anisotropy of $\gamma$ = 4.1 and approaches a two-dimensional (2D)-like behavior described by Tinkham's formula [27] (Fig. 5f). This indicates, that superconductivity appears in a locally confined region close to the interface which is in agreement with the formation of an interfacial layer showing strong $O^{2-}/As^{3-}$ imbalance.

**Discussion**

The diffusion-based design by the incorporation of impurity cations/anions into Fe-pnictide heterointerfaces provides valuable insight into the tunability of their electronic properties. The heterointerface itself (i.e. the formation of either an abrupt coherent interface or an interfacial layer) is strongly governed by electrostatic arguments.

Diffusion processes have been previously used in fabrication routes for Fe-pnictide films that make use of an ionic exchange [28, 29], and they occurred between Fe-pnictide films and substrates [30, 31]. Co-diffusion from a Ba(Fe$_{1-x}$Co$_x$)$_2$As$_2$ film into an Fe buffer layer was reported previously [30, 32]. As shown in Fig. 1a, the Ba-122/Fe heterointerface constitutes one example for a coherent heterointerface. Enhanced superconductivity in a Ba(Fe$_{1-x}$Co$_x$)$_2$As$_2$/Fe heterostructure (by more than ~3 K from ~26 to 29.4 K) was first shown in Ref. [9]. Ba-122/SrTiO$_3$ heterointerfaces (Fig. 1b,c), that depend on the surface termination and can be either abrupt (TiO$_2$ termination) or show a BaFeO$_{2+x}$ interfacial layer formation (SrO termination) [12], are also subject to interdiffusion. Although Ba(Fe$_{1.92}$Co$_{0.08}$)$_2$As$_2$/SrTiO$_3$ interfaces are atomically sharp [10], diffusion into 1.2 nm thin SrTiO$_3$ layers resulted in substituted (Ba$_{0.5}$Sr$_{0.5}$)(Ti$_{0.5}$Fe$_{0.5}$)O$_{3-x}$ and a Ti ion valence change (Ti$^{4+}$ → Ti$^{3+}$) while Sr$^{2+}$ was detected in Ba-122 [11]. Since our experiments demonstrate Co-diffusion between Fe-pnictide layers, it is questionable if a Ba-122/Ba-122 with undoped and Co-substituted layers [33] can exist. The presence of excess O$^{2-}$ in the sequential deposition of Ba(Fe$_{1.92}$Co$_{0.08}$)$_2$As$_2$ and BaFe$_2$As$_2$:O [10] resulted in nanoparticle formation rather than in abrupt Ba-122/Ba-122 interfaces as constructed in Fig. 1d, which supports our interpretation for impurity anions: Polar/polar Fe-pnictide interfaces (Figs. 1d,e) become electrostatically perturbed when O$^{2-}$ is incorporated. The replacement of pnictogen by chalcogen ions in [Fe$_2$As$_2$] changes the overall layer charge from −2 towards 0 and introduces an increasing electrostatic potential. Structural deformations like the observed faulted regions in Ba-122 (Fig. 4f,h,n,p), interfacial layer formation and the development of 2D/3D superconductivity in Sm-1111 can be correlated with it. The superconducting state in Sm-1111/Ba-122 with excess O$^{2-}$ emerges gradually from a metallic state on a time scale of several weeks to months. Induced superconductivity was previously reported for SrFe$_2$As$_2$ films, after storing them only several hours in air or water vapor [34]. There, the formation of (Fe or Sr) vacancies was suggested as crucial mechanism [35]. Another candidate would be As vacancies, that generate local magnetic moments that can coexist with superconductivity in *Ln*-1111 [36]. Although Ref. [37] mentioned the possibility to suppress the magnetic ordering (spin density wave) by intercalation of H$_2$O into SmOFeAs, we do not have yet any indication for this scenario and the true origin of the induced superconducting state is currently under investigation.

**Conclusions**

The above results indicate the significant role of ion valencies ($Co^{2+}$ and excess $O^{2-}$) for the stability as well as for the electronic ground state of Fe-pnictide heterointerfaces (*Ln*-1111/Ba-122). In the novel engineered polar/polar heterointerfaces, we find that $Co^{2+}$ preserves a coherent, epitaxial lattice matching at a commonly shared [$Fe_2As_2$] layer at the interface, whereas excess $O^{2-}$ results in the formation of an interfacial layer with a significant imbalance between $O^{2-}$ and $As^{3-}$. The appearance of superconductivity in both cases is of different origin: While a 3D superconducting state appears in the entire heterostructure due to the diffusion of $Co^{2+}$, an induced 2D/3D metal-to-superconductor transition was found upon incorporation of excess $O^{2-}$. The artificial *Ln*-1111/Ba-122 heterointerfaces offer new possibilities for tailoring electronic properties and ground states by means of ion diffusion. Questions on interface superconductivity, metal-to-superconductor transitions and the relationship between Fe-pnictide and Fe-chalcogenides can be addressed in an elegant way.

**Acknowledgments** S. H. acknowledges discussions with T. Ying. The authors thank J. Matsumoto, T. Katase, H. Hiramatsu, H. Hosono (from Tokyo Institute of Technology), as well as S. Wurmehl (from IFW Dresden) for target preparation. Mr. W.A. Smith at the University of Glasgow is thanked for his assistance with the FIB preparation, and Dr. S. McFadzean for his assistance with technical support for the JEOL ARM scanning transmission electron microscope. Drs. Damien McGrouther, Dima Maneuski and Prof. Val O'Shea are gratefully acknowledged for all their work on the development of a Medipix-3 detector into the electron microscope which became the Merlin for EM system. Dr. Gary Paterson is thanked for his assistance with providing the python libraries and example Jupyter notebooks for processing the VDF data. Dr. Stavros Nicolopoulos, Mr. Alan Robins, Mr. Doug Cosart, Dr. JK Weiss, and Dr. Jing Lu at NanoMEGAS are thanked for all their assistance with developing the prototype direct detection – SPED system, debugging the software, so that data of the quality shown here can routinely be collected. Whilst this work was not directly funded by the EPSRC, it could not have happened without the sustained investment in the development of pixelated STEM and detectors for pixelated STEM by the EPSRC over many years (funding for a broader research project on "Fast Pixel Detectors: a paradigm shift in STEM imaging" (EP/M009963/1), funding from Impact Acceleration Accounts specifically for the integration of the detector into the microscope (EP/K503903/1 & EP/R511705/1), and funding from EPSRC (EP/R511705/1) and NanoMEGAS on integration of the Merlin for EM detector with the precession diffraction system).
**Author contributions** S. H. conceived and designed the study, was responsible for thin film growth and film characterization (XRD, AFM, electrical transport measurements), evaluation of data and the preparation of the manuscript. S. N. was responsible for DFT calculations. Mi. S. was responsible for AES measurements and the quantification of the results. HAADF-STEM images and EELS data in Figs. 3 and S5 were produced by Ma. S. at Tokyo Institute of Technology. I. M. at the University of Glasgow was responsible for FIB preparation of samples for TEM/STEM analysis, and for performing the TEM and STEM analysis for scanning precession electron diffraction, EDS mapping, and atomic-resolution STEM imaging and the subsequent data analysis. All authors discussed the results.

**Competing interests** The authors declare no competing interests.

**Additional information** For the supplementary information please contact the corresponding author.

**Methods**

▪ *Engineering of thin film heterostructures and characterization* includes the growth of $LnOFe_{1-x}Co_xAs/BaFe_2As_2$ with $Ln$ = La,Sm ('$Ln$-1111/Ba-122') thin film heterostructures by pulsed laser deposition (PLD) using a Spectra Physics Quanta Ray INDI Nd:YAG(2ω) laser (λ = 532 nm, repetition rate 10 Hz, pulse width < 10 ns) in a ultra-high vacuum (UHV) chamber with ($p_{base} \approx 10^{-9} - 10^{-8}$ mbar). For the ablation $BaFe_2As_2$ and $LnO_{1-y}Fe_{1-x}Co_xAs$ targets with $x$ = 0 and 0.15 and $y$ = 0, 0.2 ($\varepsilon \approx 2 - 3$ Jcm$^{-2}$) were employed, and the $Co^{2+}$ and $O^{2-}$ supply was controlled by the $Ln$-1111 target composition. All films were deposited on MgO(100) substrates, that were pre-heat-treated in air at 800°C for 1 – 3 hours before they were inserted into the UHV environment. The substrate temperature (~ 850°C) was held constant during the deposition of both Fe-pnictide layers. Details of the film growth and target preparation are given in Refs. [26, 31, 38]. Structural characterization of the thin film heterostructures was carried out by X-ray diffraction using Bragg Brentano geometry (Rigaku Smart Lab with Cu Kα radiation with tube current of 200 mA and voltage of 45 kV). The *c*-axis lattice parameters were calculated from (00l) reflections using the extrapolation function, $f(\theta) = [\cos(\theta)\cot(\theta) + 1/\theta]$. Layer thicknesses were acquired from X-ray reflectivity (XRR) measurements (using a cross-beam optics unit and a Ge(220) monochromator). The analysis was performed using Global Fit software.

▪ *Analytical and scanning transmission electron microscopy (STEM)* were carried out for three different heterostructures.

A specimen from a $LaOFe_{1-x}Co_xAs/Ba(Fe_{1-x}Co_x)_2As_2$ heterostructure was thinned to electron transparency by ion milling at Tokyo Institute of Technology, where scanning transmission electron microscopy (STEM) was carried out on a JEOL JEM-ARM200F microscope (JEOL Ltd., Akishima, Japan) with a probe spherical aberration corrector operated at an accelerating voltage of 200 kV. The high angle annular dark field (HAADF) imaging conditions were a probe size of 0.1 nm, a semi-convergence angle of α = 32 mrad, and an annular detection angle of 80 – 170 mrad. Electron energy loss spectroscopy (EELS) was performed using a Gatan Enfina spectrometer (collection semi-angle = 12.5 mrad). The spectra were recorded with a dispersion of $\Delta E$ = 0.1 eV/channel.

Two comparable heterostructures of LaOFe$_{1-x}$Co$_x$As/Ba(Fe$_{1-x}$Co$_x$)$_2$As$_2$ and SmOFe$_{1-x}$Co$_x$As/Ba(Fe$_{1-x}$Co$_x$)$_2$As$_2$ were analyzed by HAADF-STEM at the University of Glasgow. The TEM lamellae were prepared by a standard focused ion beam (FIB) lift-out technique performed using a Xenon Plasma FIB instrument, which allowed the extraction of longer cross sections with less Ga implantation or surface damage than with a conventional Ga beam FIB [39] STEM imaging and analysis was performed using a JEOL ARM200F (scanning) transmission electron microscope equipped with a probe aberration corrector operated at 200 kV. HAADF STEM imaging and EDS were performed using a probe convergence angle of 29 mrad. HAADF STEM imaging used an inner collection angle of ~100 mrad. EDS analysis was performed using a Bruker 60 mm$^2$ SDD EDS spectrometer (Bruker Nano GmbH, Berlin, Germany) and the acquisition controlled by Spectrum Imaging within Gatan Digital Micrograph (Gatan Inc., Pleasanton, CA, USA). To allow enough signal to noise to be collected for robust and reliable detection and quantification of the results without using long dwell times per pixel and risking beam damage, the same box was scanned multiple times, with cross correlation after each scan to correct for any drift of the sample. Quantification of these results was performed by use of the *Elemental Analysis* plugin for Gatan Digital Micrograph. Scanning precession electron diffraction (SPED) was performed in transmission electron microscopy mode using the smallest condenser aperture available (10 µm) to give a low convergence angle together with a very small spot size. The precession and scanning of the beam were controlled by a NanoMEGAS DigiSTAR system (NanoMEGAS SPRL, Brussels, Belgium) with the scan areas and acquisition handled by their Topspin software. In this case, the diffraction patterns were acquired using a prototype system using a Merlin for EM direct electron detector (Quantum Detectors Ltd., Harwell, UK) instead of the normal CCD camera pointed at the focusing screen [40], as recently benchmarked and more fully described in MacLaren *et al.* [41]. Virtual dark field (VDF) imaging of the different phases was then accomplished using the method described by Paterson *et al.* [25] where the data files are read into python and processed using the *fpd* python library (https://fpdpy.gitlab.io/fpd/fpd.html) to numerically integrate the intensity in each diffraction pattern within different periodic arrays of apertures set to correspond to the diffraction spots unique to each of the crystalline phases (and avoiding diffraction spots common to more than one phase).

▪ *Auger electron spectroscopy (AES)* was carried out on an ULVAC-Phi 710 Auger electron spectrometer integrated in a scanning electron microscope (SEM) with a primary electron beam of 10 kV. An incident electron current of 10 nA on scanned areas between 9 µm² and 150 µm in diameter resulted in current densities between 57 µA·cm$^{-2}$ and 111 mA·cm$^{-2}$ on the film surfaces. For analysis clean areas without droplets were chosen. AES depth profiles were obtained by Ar$^+$ ion sputtering (1 kV on an area of 2×2 mm²). In order to decrease the effectively sputtered film area to ~0.78 mm², the film surfaces were covered by an Al foil having a hole of ~1 mm in diameter. Auger spectra, $N(E)$, were recorded in a kinetic energy range of $E$ = 30 – 1500 eV with a step size $\Delta E$ = 1 eV. The data was processed using Phi MultiPak software. The quantitative analysis of the element concentrations in the heterostructures is based on relative sensitivity factors using peak-to-peak heights of the differentiated spectra $E\cdot dN/dE$ (S5D5).

▪ *Electrical transport measurements* (four probe) measurements were carried out in a Quantum Design Physical Property Measurement System (PPMS) equipped with a sample rotator stage in external magnetic fields up to $\mu_0 H = 9$ T and down to 2 K. A constant current of 1 μA was used in all $R(T)$ and $R(H,\theta)$ measurements. For the electrical contacts Cu wires ($\varnothing = 0.01$ mm) were attached on the thin film surface using a commercial silver paste. Critical temperatures and upper critical fields, $\mu_0 H_{c2}(T,\theta)$, were evaluated using a resistivity criterion (90% or 50% $R_n$).

▪ *Electronic structure calculations* were performed within density functional theory and generalized gradient approximation [42] for the exchange correlation functional in the projector-augmented plane wave (PAW) formalism [43] as implemented in the Vienna ab-initio Simulation package (VASP) [44]. The energy cutoff was set to 500 eV, and the convergence criterion for the electronic density was chosen as $10^{-8}$ eV. For the bulk $BaFe_2As_2$ and LaFeAsO we adopted the experimental structures with the tetragonal I4/mmm and P4/nmm space groups, respectively, as reported in [45]. The La-1111/Ba-122 heterostructure used in the calculations was constructed by stacking the [001] oriented supercells with 4 $Fe_2As_2$ layers of $BaFe_2As_2$ and 5 $Fe_2As_2$ layers of LaFeAsO along the [001] axis with the Ba-As termination at the interface and the vacuum space region of 20 Å, as shown in Fig. S7. The resulting heterostructure was optimized with the force convergence criterion of $10^{-3}$ eV/Å. The Brillouin zone was sampled by a 12×12×6 mesh for the bulk $BaFe_2As_2$, a 12×12×4 mesh for the bulk LaFeAsO, and a 10×10×1 mesh for the heterostructure [46]. The calculated band structures showing the partial contributions of the Fe $d$ states for the bulk systems and the layers resolved contributions of the Fe $d$ states for the heterostructure are presented in Fig. S7.